\documentclass[aps,prd,twocolumn,superscriptaddress,nofootinbib]{revtex4-1}

\usepackage{latexsym}
\usepackage{amsmath}
\usepackage{amssymb}
\usepackage{amsfonts}
\usepackage{bm}
\usepackage{physics}

\usepackage{color}
\definecolor{purple}{rgb}{0.5,0,0.5}
\definecolor{blue}{rgb}{0.0,0,0.9}
\definecolor{prdblue}{rgb}{0.133,0.118,0.498}
\usepackage[colorlinks=true, pdfstartview=FitV, linkcolor=prdblue, citecolor= prdblue, urlcolor=prdblue]{hyperref}

\usepackage{supertabular} 
\usepackage{placeins}
\usepackage{epsfig}
\usepackage{graphicx}

\begin{document}


\title{An assessment of $\mathbf{\Upsilon}$-states above $\mathbf{B\bar B}$-threshold using a constituent-quark-model based meson-meson coupled-channels framework}


\author{P. G. Ortega}
\email[]{pgortega@usal.es}
\affiliation{Departamento de F\'isica Fundamental, Universidad de Salamanca, E-37008 Salamanca, Spain}
\affiliation{Instituto Universitario de F\'isica 
Fundamental y Matem\'aticas (IUFFyM), Universidad de Salamanca, E-37008 Salamanca, Spain}

\author{D. R. Entem}
\email[]{entem@usal.es}
\affiliation{Departamento de F\'isica Fundamental, Universidad de Salamanca, E-37008 Salamanca, Spain}
\affiliation{Instituto Universitario de F\'isica
Fundamental y Matem\'aticas (IUFFyM), Universidad de Salamanca, E-37008 Salamanca, Spain}

\author{F. Fern\'andez}
\email[]{fdz@usal.es}
\affiliation{Instituto Universitario de F\'isica 
Fundamental y Matem\'aticas (IUFFyM), Universidad de Salamanca, E-37008 Salamanca, Spain}

\author{J. Segovia}
\email[]{jsegovia@upo.es}
\affiliation{Departamento de Sistemas F\'isicos, Qu\'imicos y Naturales, Universidad Pablo de Olavide, E-41013 Sevilla, Spain}


\date{\today}

\begin{abstract}
The $\Upsilon(10753)$ state has been recently observed by the Belle and Belle~II collaborations with enough global significance to motivate an assessment of the high-energy spectrum usually predicted by any reasonable \emph{na\"ive} quark model.
In the framework of a constituent quark model which satisfactorily describes a wide range of properties of conventional hadrons containing heavy quarks, the quark-antiquark and meson-meson degrees of freedom have been incorporated with the goal of elucidating the influence of open-bottom meson-meson thresholds into the $\Upsilon$ states whose masses are within the energy range of the $\Upsilon(10753)$'s mass. It is well known that such effects could be relevant enough as to generate dynamically new states and thus provide a plausible explanation of the nature of the $\Upsilon(10753)$ state. 
In particular, we have performed a coupled-channels calculation in which the bare states $\Upsilon(4S)$, $\Upsilon(3D)$, $\Upsilon(5S)$ and $\Upsilon(4D)$ are considered together with the threshold channels $B\bar{B}$, $B\bar{B}^\ast$, $B^\ast \bar{B}^\ast$, $B_s\bar{B}_s$, $B_s\bar{B}_s^\ast$ and $B_s^\ast \bar{B}_s^\ast$. 
Among the results we have described, the following conclusions are of particular interest: (i) a richer complex spectrum is gained when thresholds are present and bare bound states are sufficiently non-relativistic; (ii) those poles obtained in the complex energy plane do not have to appear as simple peaks in the relevant cross sections; and (iii) the $\Upsilon(10750)$ candidate is interpreted as a dressed hadronic resonance whose structure is an equally mixture of a conventional $b\bar b$ state and $B^\ast \bar B^\ast$ molecule.
\end{abstract}


\maketitle


\section{INTRODUCTION}
\label{sec:intro}

The so-called $\Upsilon$-family, also known as bottomonia, are bound states made of a $b$-quark and its antiquark, $\bar{b}$, with quantum numbers $J^{PC}=1^{--}$. They were identified for the first time by the E288 Collaboration at Fermilab in 1977 while studying proton scattering on $Cu$ and $Pb$ targets in an energy regime of muon-antimuon invariant mass larger than $5\,{\rm GeV}$~\cite{E288:1977xhf, E288:1977efs}. The three observed states were called $\Upsilon(1S)$, $\Upsilon(2S)$ and $\Upsilon(3S)$; later, they were better studied at various $e^{+}e^{-}$ storage rings and through their radiative decays into the $\chi_{bJ}(2P)$ and $\chi_{bJ}(1P)$ states, with $J=0,\,1,\,2$, in a series of experiments in the 1980s~\cite{Han:1982zk, Eigen:1982zm, Klopfenstein:1983nx, Pauss:1983pa}. Despite such early experimental efforts, during the next two decades there were no significant contributions to the spectrum of the $\Upsilon$-family, except the presumably radial excitations of $S$-wave nature $\Upsilon(4S)$, $\Upsilon(10860)$ and $\Upsilon(11020)$~\cite{CLEO:1984vfn, Lovelock:1985nb}. This has been largely because the $B$-factories were not considered ideal facilities for the study of the bottomonium system since their beam energy was tuned to peak at the $\Upsilon(4S)$ mass, $10579\,\text{MeV}$, which decays in almost $100\%$ of cases to a $B\bar{B}$ pair.

The situation has changed dramatically in the last twenty years with more than two dozens of unconventional charmonium- and bottomonium-like states, the so-called XYZ mesons, observed at B-factories (BaBar, Belle and CLEO), $\tau$-charm facilities (CLEO-c and BESIII) and also proton--(anti-)proton colliders (CDF, D0, LHCb, ATLAS and CMS). For an extensive presentation of the status of heavy quarkonium physics, the reader is referred to several reviews~\cite{Lebed:2016hpi, Chen:2016qju, Chen:2016spr, Ali:2017jda, Guo:2017jvc, Olsen:2017bmm, Liu:2019zoy, Brambilla:2019esw, Yang:2020atz, Dong:2020hxe, Dong:2021bvy, Chen:2021erj, Cao:2023rhu, Mai:2022eur, Meng:2022ozq, Chen:2022asf, Guo:2022kdi, Ortega:2020tng, Huang:2023jec, Lebed:2023vnd, Zou:2021sha}.

Within the $\Upsilon$-family, the Belle collaboration~\cite{Belle:2019cbt} reported in 2019 a cross section measurement of the $e^+ e^- \to \pi^+\pi^-\Upsilon(nS)$, with $n=1,2,3$, at energies from $10.52$ to $11.02\,\text{GeV}$, observing a new structure, $\Upsilon(10753)$, with Breit-Wigner parameters:
\begin{align}
M &= \big(10752.7\pm5.9^{+0.7}_{-1.1}\big) \,\text{MeV} \,, \\
\Gamma &= \big(35.5^{+17.6}_{-11.3}{}^{+3.9}_{-3.3}\big)\,\text{MeV} \,, 
\end{align}
where the first error is statistical and the second is systematic. The global significance of the new resonance was $5.2$ standard deviations, including systematic uncertainty. Later on, the Belle~II collaboration~\cite{Belle-II:2022xdi} reported in 2022 the first observation of $\omega \chi_{bJ}(1P)$ ($J=0,1,2$) signals at $\sqrt{s}=10.745\,\text{GeV}$. By combining Belle~II data with Belle results at $\sqrt{s}=10.867\,\text{GeV}$, they find energy dependencies of the Born cross sections for $e^+e^-\to \omega\chi_{b1,b2}(1P)$ to be consistent with the shape of the $\Upsilon(10753)$ resonance; this time, the Breit-Wigner parameters were
\begin{equation}
M = \big(10753\pm6\big) \,\text{MeV} \,, \quad
\Gamma = \big(36^{+18}_{-12}\big)\,\text{MeV} \,,
\label{eq:experiment}
\end{equation}
and the suggested quantum numbers $J^P=1^-$.

Note also that experimentalists have been able to distinguish the $\Upsilon(1^{3}D_{2})$ state of the triplet $\Upsilon(1^{3}D_{J})$, with $J=1,2,3$~\cite{CLEO:2004npj, BaBar:2010tqb}. In Ref.~\cite{BaBar:2010tqb}, the $\Upsilon(1^{3}D_{2})$ was observed through the $\Upsilon(3S)\to \gamma\gamma\Upsilon(1^{3}D_{J})\to \gamma\gamma\pi^{+}\pi^{-}\Upsilon(1S)$ decay chain with a significance of $5.8\sigma$, including systematic uncertainties, and a mass of $(10164.5\pm0.8\pm0.5)\,\text{MeV}$. For the other two almost-degenerate members of the $\Upsilon(1^{3}D_{J})$ spin-triplet, $\Upsilon(1^{3}D_{1})$ and $\Upsilon(1^{3}D_{3})$, the significances were much lower, $1.8$ and $1.6$ respectively, and thus no experimental observation could be claimed.

An enormous theoretical effort has followed the experimental discoveries; in particular, focusing on the bottomonium sector, one can highlight the work done using Lattice-regularized QCD~\cite{Ryan:2020iog, Bulava:2022ovd}, functional methods~\cite{Hilger:2014nma, Fischer:2014cfa, Yin:2019bxe, Yin:2021uom}, QCD sum rules~\cite{Parui:2021jbu, Chen:2013tma, Kleiv:2013fhe}, effective field theories~\cite{Brambilla:2004jw, Brambilla:1999xf, Segovia:2018qzb, Peset:2018jkf} and quark models~\cite{Barnes:1996ff, Ebert:2002pp, Eichten:2007qx, Ferretti:2013vua, Godfrey:2015dia}. Most of the mentioned references focus on the description of conventional bottomonium. This is because the first open-bottom threshold is higher in energy than the corresponding one in the charmonium sector and thus a larger number of conventional states are expected below $B\bar B$-threshold. Moreover, the only experimentally discovered excited states which are above the $B\bar B$-threshold, and so they have the ability to be strongly influenced by meson-meson thresholds, are $\Upsilon(4S)$, $\Upsilon(10860)$ and $\Upsilon(11020)$, besides the unconfirmed state $\Upsilon(10753)$.

Herein, we study the high-energy spectrum of the $\Upsilon$-family in the framework of a constituent quark model~\cite{Vijande:2004he} which satisfactorily describes a wide range of properties of conventional hadrons containing heavy quarks~\cite{Segovia:2013wma, Segovia:2016xqb}. The quark-antiquark and meson-meson degrees of freedom are incorporated with the goal of elucidating the influence of open-bottom meson-meson thresholds in the conventional states but, above all, to shed some light on the nature and structure of the $\Upsilon(10753)$ state. We should briefly mention that \emph{charged} bottomonium-like states $Z_b(10610)^\pm$ and $Z_b(10650)^\pm$ were identified by the Belle Collaboration~\cite{Belle:2011aa} as peaks in the invariant mass distribution of the $\pi^\pm h_b(mP)$ ($m=1,2$) and $\pi^\pm \Upsilon(nS)$ ($n=1,2,3$) subsystems when the $\Upsilon(10860)$ resonance decays into two pions plus an $h_b$ or $\Upsilon$. The quantum numbers of the $Z_b$'s were analyzed to be $I^G(J^{PC})=1^+(1^{+-})$~\cite{Adachi:2011mks} and so they belong to the isospin $I=1$ sector of bottomonium-like particles, disconnected from the conventional bottomonium states of isospin $I=0$. In fact such exotic mesons were studied by us in Ref.~\cite{Ortega:2021xst}.

A variational formalism based on a highly efficient numerical approach named the Gaussian expansion method (GEM)~\cite{Hiyama:2003cu} is used to solve the bottomonium Hamiltonian. Moreover, this Gaussian expansion allows us to compute effective meson-meson interactions from the original quark--(anti-)quark potentials in a simplified way through the so-called Resonating Group Method~\cite{Wheeler:1937zza, Tang:1978zz}. Finally, within our approach, the coupling between the quark-antiquark and meson-meson sectors requires the creation of a light quark-antiquark pair. Therefore, the  associated operator should be similar to the one describing open-flavor meson strong decays and we adopt the $^{3}P_{0}$ transition operator described in, for instance, Ref.~\cite{Micu:1968mk}. This theoretical formalism has the advantage of easily introducing the coupling with all meson-meson partial waves and the straightforward computation of the probabilities associated with the different Fock components of the physical state.

The manuscript is organized as follows. After this introduction, the theoretical framework is briefly presented in section~\ref{sec:theory}. Section~\ref{sec:results} is mainly devoted to the analysis and discussion of our theoretical results. Finally, we summarize and draw some conclusions in Sec.~\ref{sec:summary}.


\section{THEORETICAL FRAMEWORK}
\label{sec:theory}

\subsection{CONSTITUENT QUARK MODEL}
\label{subsec:CQM}

Among the wide range of chiral quark models developed in the last 50 years~\cite{Fernandez:2021zjq}, our theoretical framework is a QCD-inspired constituent quark model (CQM) proposed in Ref.~\cite{Vijande:2004he} and extensively reviewed in Refs.~\cite{Segovia:2013wma, Segovia:2016xqb}. Moreover, the CQM has been recently applied with success to conventional mesons containing heavy quarks, describing a wide range of physical observables that concern spectra~\cite{Segovia:2008zz, Segovia:2015dia, Ortega:2016hde, Ortega:2020uvc}, strong decays~\cite{Segovia:2012cd, Segovia:2009zz, Segovia:2011zza, Segovia:2013kg}, hadronic transitions~\cite{Segovia:2014mca, Segovia:2015raa, Martin-Gonzalez:2022qwd} as well as electromagnetic and weak reactions~\cite{Segovia:2011dg, Segovia:2012yh, Segovia:2013sxa}.

The main pieces of our CQM are spontaneous chiral symmetry breaking of the QCD Lagrangian together with perturbative one-gluon exchange (OGE) and non-perturbative color confining interactions. In the heavy quark sector, chiral symmetry is explicitly broken and Goldstone-boson exchanges do not appear. Thus, OGE and confinement are the only interactions remaining. 

The OGE potential contains central, tensor and spin-orbit contributions given by
\begin{widetext}
\begin{subequations}
\begin{align}
V_{\rm OGE}^{\rm C}(\vec{r}_{ij}) &= +
\frac{1}{4}\alpha_{s}(\vec{\lambda}_{i}^{c}\cdot
\vec{\lambda}_{j}^{c})\Bigg[ \frac{1}{r_{ij}} - \frac{1}{6m_{i}m_{j}} (\vec{\sigma}_{i}\cdot\vec{\sigma}_{j}) 
\frac{e^{-r_{ij}/r_{0}(\mu)}}{r_{ij}r_{0}^{2}(\mu)} \Bigg] \,, \\
V_{\rm OGE}^{\rm T}(\vec{r}_{ij}) &= -\frac{1}{16}\frac{\alpha_{s}}{m_{i}m_{j}}
(\vec{\lambda}_{i}^{c}\cdot\vec{\lambda}_{j}^{c}) \Bigg[ 
\frac{1}{r_{ij}^{3}} - \frac{e^{-r_{ij}/r_{g}(\mu)}}{r_{ij}}\left( 
\frac{1}{r_{ij}^{2}}+\frac{1}{3r_{g}^{2}(\mu)}+\frac{1}{r_{ij}r_{g}(\mu)}\right)
\Bigg] S_{ij} \,, \\
V_{\rm OGE}^{\rm SO}(\vec{r}_{ij}) &=
-\frac{1}{16}\frac{\alpha_{s}}{m_{i}^{2}m_{j}^{2}}(\vec{\lambda}_{i}^{c} \cdot
\vec{\lambda}_{j}^{c}) \Bigg[ \frac{1}{r_{ij}^{3}}-\frac{e^{-r_{ij}/r_{g}(\mu)}}
{r_{ij}^{3}} \left(1+\frac{r_{ij}}{r_{g}(\mu)}\right) \Bigg] \nonumber \\
&
\times \left[((m_{i}+m_{j})^{2}+2m_{i}m_{j})(\vec{S}_{+}\cdot\vec{L})+
(m_{j}^{2}-m_{i}^{2}) (\vec{S}_{-}\cdot\vec{L}) \right] \,,
\end{align}
\label{eq:OGEpot}
\end{subequations}
\end{widetext}
where $r_{0}(\mu)=\hat{r}_{0}\frac{\mu_{nn}}{\mu_{ij}}$ and
$r_{g}(\mu)=\hat{r}_{g}\frac{\mu_{nn}}{\mu_{ij}}$ are regulators which depend on
$\mu_{ij}$, the reduced mass of the $q\bar{q}$-pair; for example, $\mu_{nn}=m_n/2$ with $m_n$ the mass of the light quark with $n=u$- or $d$-quark. The contact term of the central potential has been regularized as follows
\begin{equation}
\delta(\vec{r}_{ij})\approx \frac{1}{4\pi 
r_{0}^{2}}\frac{e^{-r_{ij}/r_{0}}}{r_{ij}} \,.
\label{eq:delta}
\end{equation}

The wide energy range needed to provide a consistent description of light, strange and heavy mesons requires an effective scale-dependent strong coupling constant. We use the frozen coupling constant of Ref.~\cite{Vijande:2004he},
\begin{equation}
\alpha_{s}(\mu_{ij})=\frac{\alpha_{0}}{\ln\left( 
\frac{\mu_{ij}^{2}+\mu_{0}^{2}}{\Lambda_{0}^{2}} \right)} \,,
\end{equation}
in which $\alpha_{0}$, $\mu_{0}$ and $\Lambda_{0}$ are parameters of the model determined by a global fit to the meson spectra.

The different pieces of the color confining potential are
\begin{subequations}
\begin{align}
V_{\rm CON}^{\rm C}(\vec{r}_{ij}) &= \left[-a_{c}(1-e^{-\mu_{c}r_{ij}})+\Delta
\right] (\vec{\lambda}_{i}^{c}\cdot\vec{\lambda}_{j}^{c}) \,, \\
V_{\rm CON}^{\rm SO}(\vec{r}_{ij}) &=
-(\vec{\lambda}_{i}^{c}\cdot\vec{\lambda}_{j}^{c}) \frac{a_{c}\mu_{c}e^{-\mu_{c}
r_{ij}}}{4m_{i}^{2}m_{j}^{2}r_{ij}} \nonumber \\
&
\times  \Bigg[((m_{i}^{2}+m_{j}^{2})(1-2a_{s}) \nonumber \\
&
\quad\,\, + 4m_{i}m_{j}(1-a_{s}))(\vec{S}_{+} \cdot\vec{L}) \nonumber \\
&
\quad\,\, +(m_{j}^{2}-m_{i}^{2}) (1-2a_{s}) (\vec{S}_{-}\cdot\vec{L}) \Bigg] \,,
\end{align}
\end{subequations}
where the mixture between scalar and vector Lorentz structures of the color confinement is controlled by $a_{s}$. Besides, this potential presents at short inter-quark distances a linear behavior with an effective confinement strength given by $\sigma = -a_{c}\,\mu_{c}\,(\vec{\lambda}^{c}_{i}\cdot \vec{\lambda}^{c}_{j})$, while it becomes constant at very large inter-quark distances showing a threshold defined by
\begin{equation}
V_{\text{thr}} = (-a_{c}+ \Delta) \cdot (\vec{\lambda}^{c}_{i}\cdot
\vec{\lambda}^{c}_{j}) \,;
\end{equation}
\emph{viz.} there is no $q\bar{q}$ bound states with eigenenergies higher than this threshold; the system suffers a transition from a color string configuration between two static color sources into a pair of static mesons due to the breaking of the color string and the most favored decay into hadrons.

Among the different theoretical formalisms to solve the Schr\"odinger equation, in order to find the quark-antiquark bound states, we use the Rayleigh-Ritz variational method in which the wave function solution of the Schr\"odinger equation is expanded as indicated by the Gaussian Expansion Method (GEM)~\cite{Hiyama:2003cu}. It has also been proven to be quite efficient on solving the bound-state problem of a few-body system~\cite{Yang:2017qan, Yang:2019lsg, Yang:2020fou, Yang:2021zhe, Yang:2020twg, Yang:2018oqd}, providing enough accuracy and simplifying the evaluation of matrix elements.

The radial wave function is then expressed as
\begin{equation}
R_{\alpha}(r) = \sum_{n=1}^{n_{max}} c_{n}^\alpha \phi^G_{nl}(r) \,,
\end{equation} 
where $\alpha$ refers to the channel quantum numbers. The coefficients, $c_{n}^\alpha$, and the eigenvalue, $E$, are determined from the Rayleigh-Ritz variational principle:
\begin{equation}
\sum_{n'=1}^{n_{max}} \left[\left(T_{nn'}^\alpha - E N_{nn'}^\alpha\right)
c_{n'}^\alpha + \sum_{\alpha'=1}^{\text{channels}} \ V_{nn'}^{\alpha\alpha'} c_{n'}^{\alpha'} = 0\right] \,,
\end{equation}
where $T_{nn'}^\alpha$, $N_{nn'}^\alpha$ and $V_{nn'}^{\alpha\alpha'}$ are, respectively, the matrix elements of the kinetic energy, the normalization and the potential. $T_{nn'}^\alpha$ and $N_{nn'}^\alpha$ are diagonal, whereas the mixing between different channels is given by $V_{nn'}^{\alpha\alpha'}$.

Following Ref.~\cite{Hiyama:2003cu}, we employ Gaussian trial functions with
ranges in geometric progression. This enables the optimization of ranges
employing a small number of free parameters. Moreover, the geometric
progression is dense at short distances, so that it enables the description of
the dynamics mediated by short range potentials. The fast damping of the
Gaussian tail does not represent an issue, since we can choose the maximal
range much longer than the hadronic size.


\subsection{COUPLED-CHANNELS CALCULATION}
\label{subsec:coupledchannel} 

It is well known that conventional mesons must be influenced in a non-perturbative way by meson-meson thresholds when these are close. In order to take into account this effect within the bottomonium sector, we perform a coupled-channels calculation in which the total hadron wavefunction is described by a combination of conventional $b\bar b$ states and open-bottom meson-meson channels:
\begin{equation}
| \Psi \rangle = \sum_\alpha c_\alpha | \psi_\alpha \rangle
+ \sum_\beta \chi_\beta(P) |\phi_A \phi_B \beta \rangle \,,
\label{ec:funonda}
\end{equation}
where $|\psi_\alpha\rangle$ are $b\bar{b}$ eigenstates of the two-body Hamiltonian, $\phi_{M}$ is the wavefunction associated with meson $M=A, B$, $|\phi_A \phi_B \beta \rangle$ is the two meson state whose quantum numbers are $\beta$ and $\chi_\beta(P)$ is the relative wave function between the two mesons.

Under the above assumption, two sources of interaction emerge. On one hand, the two- and four-quark sectors can be coupled via the creation of a light-quark-antiquark pair. On the other hand, there is a residual interaction, derived from the microscopic quark--(anti-)quark potential described by the CQM, among the two mesons. To derive the latter, we use the Resonating Group Method (RGM)~\cite{Wheeler:1937zza, Tang:1978zz} (see also Refs.~\cite{Ortega:2012rs, Ortega:2020tng} for further details).

Within RGM, the wave function of a system composed of two mesons with distinguishable quarks is given by
\begin{equation}
\langle \vec{p}_{A} \vec{p}_{B} \vec{P} \vec{P}_{\rm c.m.} | \phi_A \phi_B \beta
\rangle = \phi_{A}(\vec{p}_{A}) \phi_{B}(\vec{p}_{B})
\chi_{\beta}(\vec{P}) \,,
\label{eq:wf}
\end{equation}
where, \emph{e.g.}, $\phi_{A}(\vec{p}_{A})$ is the wave function of the meson $A$ with $\vec{p}_{A}$ the relative momentum between its quark and antiquark. The wave function $\chi_\beta(\vec{P})$ takes into account the relative motion between the two mesons.

A general process $AB\to A'B'$ can be described by means of either direct or exchange potentials; the last ones appear due to the possibility of having to consider quark exchanges between clusters. In this study we do not have this case, so we only have direct potentials, which can be written as
\begin{align}
&
{}^{\rm RGM}V_{D}^{\beta\beta '}(\vec{P}',\vec{P}) = \sum_{i\in A, j\in B} \int d\vec{p}_{A'} d\vec{p}_{B'} d\vec{p}_{A} d\vec{p}_{B} \nonumber \\
&
\times \phi_{A}^{\ast}(\vec{p}_{A'}) \phi_{B}^{\ast}(\vec{p}_{B'})
V_{ij}^{\beta\beta '}(\vec{P}',\vec{P}) \phi_{A'}(\vec{p}_{A}) \phi_{B'}(\vec{p}_{B})  \,.
\label{eq:RGMdir}
\end{align}
where $\beta^{(\prime)}$ labels the set of quantum numbers needed to uniquely define a certain meson-meson partial wave, $\vec P^{(\prime)}$ are the initial (final) relative momentum of the meson-meson pair, and $V_{ij}^{\beta\beta '}(\vec{P}',\vec{P})$ are the microscopic quark--(anti-)quark potentials from the CQM and the sum runs over the constituent particles inside each meson cluster.

The coupling between the bottomonia and the open-bottom meson-meson thresholds requires the creation of a light quark-antiquark pair. For that purpose, we use the $^{3}P_{0}$ transition operator which was originally introduced in the 1970s to describe strong decays of mesons and baryons~\cite{Micu:1968mk, LeYaouanc:1972vsx, LeYaouanc:1973ldf}. The associated non-relativistic operator can be written as~\cite{Segovia:2012cd, Segovia:2013kg}:
\begin{align}
T &= -\sqrt{3} \, \sum_{\mu,\nu}\int d^{3}\!p_{\mu}d^{3}\!p_{\nu}
\delta^{(3)}(\vec{p}_{\mu}+\vec{p}_{\nu})\frac{g_{s}}{2m_{\mu}}\sqrt{2^{5}\pi} \nonumber \\
&
\times \left[\mathcal{Y}_{1}\left(\frac{\vec{p}_{\mu}-\vec{p}_{\nu}}{2}
\right)\otimes\left(\frac{1}{2}\frac{1}{2}\right)1\right]_{0}a^{\dagger}_{\mu}
(\vec{p}_{\mu})b^{\dagger}_{\nu}(\vec{p}_{\nu}) \,.
\label{eq:Otransition2}
\end{align}
where $\mu$ $(\nu)$ are the spin, flavor and color quantum numbers of the created quark (antiquark). The orbital angular momentum and spin of the pair are both equal to one. Note that ${\cal Y}_{lm}(\vec{p}\,) = p^{l} Y_{lm}(\hat{p})$ is the solid harmonic defined in function of the spherical one. The unique parameter of the decay model is the strength of the quark-antiquark pair creation from the vacuum, $\gamma=g_{s}/2m$, where $m$ is the mass of the created quark (antiquark).

The values of $\gamma$ can be constrained through meson strong decays. A global fit to charmed, charmed-strange, hidden-charm and hidden-bottom sectors was performed in Ref.~\cite{Segovia:2012cd}, finding a running of the strength parameter given by
\begin{equation}
\gamma(\mu) = \frac{\gamma_{0}}{\log\left(\frac{\mu}{\mu_{0}}\right)} \,,
\label{eq:fitgamma}
\end{equation}
where $\gamma_{0}$ and $\mu_{0}$ are free parameters, whereas $\mu$ is the reduced mass of the constituent quark-antiquark pair of the decaying meson. In this work, we use the value of $\gamma$ corresponding to the bottomonium sector, \emph{i.e.} $\gamma=0.205$.

From the operator in Eq.~(\ref{eq:Otransition2}), we define the meson to meson-meson transition potential $h_{\beta \alpha}(P)$ as
\begin{equation}
\langle \phi_{M_1} \phi_{M_2} \beta | T | \psi_\alpha \rangle =
\delta^{(3)}(\vec P_{\rm cm})\,P \, h_{\beta \alpha}(P),
\label{Vab}
\end{equation}
where $P$ is the relative momentum of the two-meson state. In order to soften the $^3P_0$ production vertex at high momenta, we follow the suggestion of Ref.~\cite{Morel:2002vk} and used a Gaussian-like momentum-dependent form factor to truncate the vertex,
\begin{equation}
h_{\beta \alpha}(P) \,\to h_{\beta \alpha}(P) e^{-\frac{P^2}{2\Lambda^2}} \,,
\end{equation}
with $\Lambda=0.84\,{\rm GeV}$. This cut-off's value is taken from similar analysis~\cite{Ortega:2016mms, Ortega:2016pgg}, so no fine-tuning of parameters is employed in the present work.

Finally, within the formalism explained above, the coupled-channels
equations can be written as
\begin{align}
&
c_\alpha M_\alpha +  \sum_\beta \int h_{\alpha\beta}(P) \chi_\beta(P)P^2 dP = E
c_\alpha\,, \\
&
\sum_{\beta}\int H_{\beta'\beta}(P',P)\chi_{\beta}(P) P^2 dP + \nonumber \\
&
\hspace{2.50cm} + \sum_\alpha h_{\beta'\alpha}(P') c_\alpha = E
\chi_{\beta'}(P')\,,
\label{ec:Ec-Res}
\end{align}
where $M_\alpha$ are the masses of the bare $b\bar{b}$ mesons and 
$H_{\beta'\beta}$ is the RGM Hamiltonian for the two-meson states obtained from
the naive CQM interaction. In order to explore states above and below meson-meson thresholds, the former coupled-channels equations should then be written as a set of coupled Lippmann-Schwinger equations,
\begin{align}
T_{\beta}^{\beta'}(E;p',p) &= V_{\beta}^{\beta'}(p',p) + \sum_{\beta''} \int
dp''\, p^{\prime\prime2}\, V_{\beta''}^{\beta'}(p',p'') \nonumber \\
&
\times \frac{1}{E-{\cal E}_{\beta''}(p^{''})}\, T_{\beta}^{\beta''}(E;p'',p) \,,
\end{align}
where $V_{\beta}^{\beta'}(p',p)$ is the projected potential that contains the sum of direct potentials, Eq.~\eqref{eq:RGMdir}, as well as the effective potential which encodes the coupling with the two-quark sector:
\begin{equation}
\label{eq:effectiveV}
V_{\rm eff}^{\beta'\beta}(P',P;E)=\sum_{\alpha}\frac{h_{\beta'\alpha}(P')
h_{\alpha\beta}(P)}{E-M_{\alpha}} \,.
\end {equation}
The ${\cal E}_{\beta''}(p'')$ is the energy corresponding to a momentum $p''$, which in the nonrelativistic case is
\begin{equation}
{\cal E}_{\beta}(p) = \frac{p^2}{2\mu_{\beta}} + \Delta M_{\beta} \,.
\end{equation}
Herein, $\mu_{\beta}$ is the reduced mass of the meson-meson system corresponding to the channel $\beta$, and $\Delta M_{\beta}$ is the difference between the meson-meson threshold and the lightest one, taken as a reference.

Once the $T$-matrix is calculated, we determine its on-shell part which is directly related to the scattering matrix in the case of non-relativistic kinematics:
\begin{equation}
S_{\beta}^{\beta'} = 1 - 2\pi i
\sqrt{\mu_{\beta}\mu_{\beta'}k_{\beta}k_{\beta'}} \,
T_{\beta}^{\beta'}(E+i0^{+};k_{\beta'},k_{\beta}) \,,
\end{equation}
with $k_{\beta}$ the on-shell momentum for channel $\beta$. All the potentials shall be analytically continued for complex momenta; this allows us to find the poles of the $T$-matrix in any possible Riemann sheet.


\begin{figure}[!t]
\centerline{\includegraphics[width=0.45\textwidth]{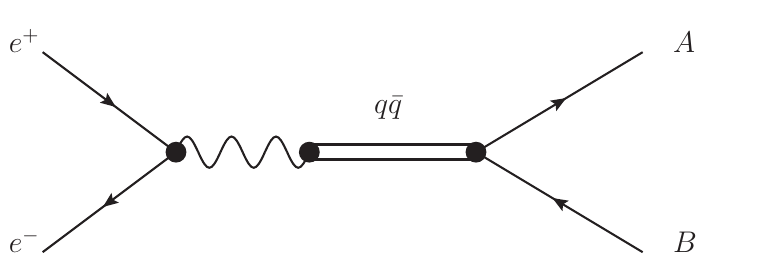}}
\caption{\label{fig:dia} Production of open-bottom mesons denoted by the $AB$-pair through a $q\bar q$ resonance with $J^{PC}=1^{--}$. In this case a conventional state of the $\Upsilon$-family is produced that decays later on in a pair of open-bottom mesons.}
\end{figure}

\subsection{Production in annihilation $e^+e^-$ through a resonance}
\label{subsec:epempro}

For later convenience, our objective herein is to calculate cross section of the process $e^+e^- \to AB$ (represented in Fig.~\ref{fig:dia}), with $A$ or $B$ denoting open-bottom mesons, through an arbitrary set of $b\bar b$ resonances with quantum numbers $J^{PC}=1^{--}$. For that purpose we use the extended Van Royen-Weisskopf formalism for meson leptonic decays, considering non-zero momentum distribution for the quark (antiquark) inside the meson~\cite{Blanco:2001ip}. It should be noticed that the Feynman diagram in Fig.~\ref{fig:dia} can be used when the quark-antiquark pair acts as free particles and has a certain momentum, in this case weighted with the corresponding $\Upsilon$ meson's wave function, $\phi(\vec{p})$, which gives the probability amplitude of finding a quark with momentum $\vec{p}$ inside the $b\bar b$ meson.

The process to be studied is $\langle e^+e^-|AB \rangle_\beta$ through one or few $|q\bar q\rangle_\alpha$ resonances. This process can be factorized as,
\begin{equation}
\langle e^+e^-|AB \rangle_\beta =\sum_\alpha \langle e^+e^-|q\bar q\rangle_\alpha \,_\alpha\langle q\bar q|AB\rangle_\beta \,.
\end{equation}

In the center-of-mass reference system, the $S$-matrix of the process $\langle e^+e^-|q\bar q\rangle_\alpha$ can be written as
\begin{align}
&
S = -ie^2e_q (2\pi)^4 \int d^3p\frac{ \delta^{(4)}(P_i-P_f)m_l m_q}{(2\pi)^3 E_p (2\pi)^3 E_q} \nonumber \\
&
\times \sum_{M_1 M_2\mu_L\mu_S} \left< L\mu_LS\mu_S|J\mu_J\right> \langle \frac{1}{2}M_1\frac{1}{2}M_2|S\mu_S \rangle \phi(\vec p)\frac{g_{\mu\nu}}{s} \nonumber \\
& 
\times \left[ \bar u_l(q,\xi_1)\gamma^\mu v_l(-q,\xi_2) \right]    \left[ \bar v_q(-p,M_2) \gamma^\nu u_q(p,M_1) \right] \,,
\end{align}
where $\{m_l,e,E_q,\vec q,\xi\}$ represents the mass, charge, energy, momentum and spin projection of the incoming electron (positron); $\{m_q,e_q,E_p,\vec p,M\}$ the same for the created quark (antiquark), bound in a $J^{PC}=1^{--}$ state with quantum numbers $\{J,L,S\}$ and $\{\mu_J,\mu_L,\mu_S\}$ projections. We express the Clebsch-Gordan coefficients as $\langle j_1m_1\,j_2m_2|j_3m_3\rangle$.
The virtual photon's four-momentum verifies $k_\gamma^2=s$.

The above $S$-matrix expression can be simplified and written in terms of the invariant amplitude ${\cal M}$,

\begin{equation}
S = -2\pi i\delta^{(4)}(\sum p_f -\sum p_i) \mathcal{M} \,,
\end{equation}
so we arrive at
\begin{align}
\mathcal{M} &= e^2 e_q \frac{1}{(2\pi)^{3/2}} \frac{m_l}{E_q}  \frac{2}{3s} \overline{\Psi(0)} (-1)^{1/2+\xi_2} \frac{E_q+m_l}{2m_l} \nonumber \\
&
\Bigg[ \left(1+\frac{q^2}{(E_q+m_l)^2}\right) \langle \frac{1}{2}\xi_1\frac{1}{2}\xi_2|1\mu_J \rangle \nonumber \\
&
- \frac{2}{(E_q+m_l)^2}\sum_n (-1)^n \langle \frac{1}{2}\xi_1\frac{1}{2}\xi_2|1n \rangle q_{-n}q_{\mu_J} \Bigg] \,,
\end{align}
where we have defined
\begin{equation}
\overline{\Psi(0)} = \left[ \Psi(0) \delta_{L,0} - \frac{\sqrt{2}}{6\pi} \sqrt{2L+1} \langle L010|10 \rangle I_4 \right] \,.
\end{equation}
Here, $\Psi(0)$ is the $\Upsilon$ meson wave function at the origin. The second term
encodes the contribution of $L=2$ states, where
\begin{equation}\label{eq:I4integral}
I_4 \equiv \int \frac{p^4 dp R_r(p)}{E_p(E_p+m_q)} \,.
\end{equation}
It is worth noticing that this definition of $\overline{\Psi(0)}$ allows the contribution of both $\Upsilon$ $^3S_1$  and $\Upsilon$ $^3D_1$ states due to the non-zero quark momentum distribution inside the meson.

Concerning the decay of the $\Upsilon$ meson into two mesons, we can extract the  $_\alpha\langle q\bar q|AB \rangle_\beta$ amplitude from the coupled channels formalism~\cite{Ortega:2012rs}, and it is given by
\begin{equation}
_\alpha\langle q\bar q|AB\rangle_\beta = \sum_{\alpha'} \bar h_{\alpha'\beta}(\sqrt{s};k) \Delta_{\alpha'\alpha}(\sqrt{s})^{-1}  \,,
\end{equation}
with $k$ the relative on-shell momentum of the two mesons and
$\Delta_{\alpha'\alpha}$ the full resonance propagator, given by

\begin{align}
 \Delta_{\alpha'\alpha}(E)=(E-M_{\alpha})\delta_{\alpha'\alpha}+\mathcal{
G}_{\alpha'\alpha}(E)
\end{align}
with $\mathcal{G}_{\alpha'\alpha}$ the mass-shift function

\begin{align}
\mathcal{G}_{\alpha'\alpha}(E)=\sum_\beta \int \dfrac{\bar h_{\alpha\beta}(E;q)h_{\beta \alpha'}(q)}{q^2/2\mu+m_A+m_B-E}q^2 dq\,.
\end{align}

The function $\bar h^{\beta\alpha}$ can be interpreted as the $^3P_0$ vertex
dressed by the meson-meson interaction
\begin{equation}
\bar h_{\beta\alpha}(E;P)= h_{\beta\alpha}(P)-\sum_{\beta'}\int
\frac{T^{\beta\beta'}_V(P,q;E)h_{\beta'\alpha}(q)}
{q^2/2\mu+m_A+m_B-E}\,q^2dq\,,
\end{equation}
where $T^{\beta'\beta}_V(P',P;E)$ is the $T$ matrix of the RGM potential excluding the
coupling to the $q\bar q$ pairs.

The expression for the total cross section $e^+e^-\to b\bar b \to AB$ in the center-of-mass reference system is given by
\begin{equation}
d\sigma_\beta= (2\pi)^4 \frac{E_AE_B}{\sqrt{s}k_0}\delta(k-k_0)\frac{E_q}{2|q|}|\mathcal{M}_\beta|^2 d^3k \,,
\end{equation}
where the on-shell momentum is
\begin{equation}
k_o=\frac{\sqrt{\left[s-(m_A+m_B)^2\right]\left[s-(m_A-m_B)^2\right]}}{2\sqrt{s}} \,.
\end{equation}

Averaging ${\cal M}$ over the polarizations of the initial states and sum over final states, we arrive at
\begin{align}
\sigma_\beta &= \frac{4\pi^2}{3} e^4 e_q^2\frac{\sqrt{k_o^2+m_A^2}\sqrt{k_o^2+m_B^2}k_o}{s^{5/2}} \nonumber \\
&
\times \Big| \sum_{\nu,\nu'} \phi_{\nu'\beta}(k_o;\sqrt{s}) \Delta_{\nu'\nu}(\sqrt{s})^{-1} \overline{\Psi_\nu(0)} \Big|^2 \,,
\end{align}
which only depends on the on-shell momentum of the mesons in the final state.


\begin{table}[!t]
\caption{\label{tab:parameters} Quark model parameters.}
\begin{ruledtabular}
\begin{tabular}{lrr}
Quark masses & $m_{n}$ (MeV) & $313$ \\
             & $m_{s}$ (MeV) & $555$ \\
             & $m_{b}$ (MeV) & $5110$ \\
\hline
OGE & $\hat{r}_{0}$ (fm) & $0.181$ \\
    & $\hat{r}_{g}$ (fm) & $0.259$ \\
    & $\alpha_{0}$ & $2.118$ \\
    & $\Lambda_{0}$ $(\mbox{fm}^{-1})$ & $0.113$ \\
    & $\mu_{0}$ (MeV) & $36.976$ \\
\hline
Confinement & $a_{c}$ (MeV) & $507.4$ \\
            & $\mu_{c}$ $(\mbox{fm}^{-1})$ & $0.576$ \\
            & $\Delta$ (MeV) & $184.432$ \\
            & $a_{s}$ & $0.81$ \\
\end{tabular}
\end{ruledtabular}
\end{table}

\begin{table}[!t]
\caption{\label{tab:predmassesbb} Masses, in MeV, of the bottomonium states with quantum numbers $J^{PC}=1^{--}$ (the so-called $\Upsilon$-family) predicted by our constituent quark model. $nL$ identifies the radial excitation, being $n=1$ the ground state, and the dominant orbital angular momentum component in each wave function.}
\begin{ruledtabular}
\begin{tabular}{ccccc}
State & $J^{PC}$ & $nL$ & The. (MeV) & Exp. (MeV) \cite{ParticleDataGroup:2022pth} \\
\hline
$\Upsilon$     & $1^{--}$ & $1S$ & $9502$  & $9460.40\pm0.10$ \\
               &          & $2S$ & $10015$ & $10023.4\pm0.5$ \\
               &          & $1D$ & $10117$ & - \\
               &          & $3S$ & $10349$ & $10355.1\pm0.5$ \\ 
               &          & $2D$ & $10414$ & - \\
               &          & $4S$ & $10607$ & $10579.4\pm1.2$ \\
               &          & $3D$ & $10653$ & - \\
               &          & $5S$ & $10818$ & $10885.2\pm3.1$ \\
               &          & $4D$ & $10853$ & - \\
               &          & $6S$ & $10995$ & $11000\pm4$ \\
               &          & $5D$ & $11023$ & - \\
\end{tabular}
\end{ruledtabular}
\end{table}

\section{RESULTS}
\label{sec:results}

The relevant parameters of our \emph{na\"ive} CQM are shown in Table~\ref{tab:parameters}, and they are the same used in, \emph{e.g.}, Ref.~\cite{Segovia:2016xqb}. Table~\ref{tab:predmassesbb} shows the predicted bottomonium states with quantum numbers $J^{PC}=1^{--}$ as well as the world average masses reported in the Review of Particle Physics (RPP)~\cite{ParticleDataGroup:2022pth} provided by the Particle Data Group. It is inferred from Table~\ref{tab:predmassesbb} that a global description of the $\Upsilon$-family is obtained by our CQM. 

It is also evident that the model does not provide a conventional state compatible with the experimentally observed $\Upsilon(10753)$ state. One would be tempted to assert that our CQM has some major deficiency; however, it is well known that any reasonable quark model describes well the bottomonium sector, providing a comparable spectrum and having very similar characteristics. This fact may imply that the $\Upsilon(10753)$ cannot be explained as a conventional bottomonium system and thus it has an exotic origin. In other words, there should be another important mechanism in the dynamics of the $\Upsilon(10753)$ that is not implemented in our, even any, \emph{na\"ive} CQM.

\begin{table}[!t]
\caption{\label{tab:thresholds} Masses, in MeV, of the isospin-averaged $B\bar{B}$, $B\bar{B}^\ast$, $B^\ast \bar{B}^\ast$, $B_s\bar{B}_s$, $B_s\bar{B}_s^\ast$ and $B_s^\ast \bar{B}_s^\ast$ thresholds, from PDG~\cite{ParticleDataGroup:2022pth}.}
\begin{ruledtabular}
\begin{tabular}{ccccccc}
Channel & $B\bar{B}$ & $B\bar{B}^\ast$ & $B^\ast \bar{B}^\ast$ & $B_s\bar{B}_s$ & $B_s\bar{B}_s^\ast$ & $B_s^\ast \bar{B}_s^\ast$ \\ 
\hline
Mass & $10558.8$ & $10604.1$ & $10649.4$ & $10733.8$ & $10782.7$ & $10831.6$ \\ 
\end{tabular}
\end{ruledtabular}
\end{table}

The lowest open-bottom meson-meson threshold is $B\bar B$, with a non-interacting mass of about $10.56\,\text{GeV}$. Moreover, the dominant open-bottom meson-meson strong decay channels of the $\Upsilon$-family are considered to be $B\bar{B}$, $B\bar{B}^\ast$, $B^\ast \bar{B}^\ast$, $B_s\bar{B}_s$, $B_s\bar{B}_s^\ast$ and $B_s^\ast \bar{B}_s^\ast$~\footnote{For now on, we denote $B_{(s)}\bar B^*_{(s)}\equiv B_{(s)}\bar B^*_{(s)}+$h.c.}. Their non-interacting mass thresholds are shown in Table~\ref{tab:thresholds} and they belong to an energy range between $10.56$ to $10.83\,\text{GeV}$. In order to assess agreement between theory and experiment, we should include those coupled-channels effects that may play an important role in our description of the $\Upsilon$-family, at least, for the $\Upsilon(4S)$ up to the $\Upsilon(4D)$ which fall within the range of energies of the most important open-flavor meson-meson thresholds.

The coupling of bare $b\bar b$ states with open-bottom meson-meson channels depends on the relative position of the $b\bar b$ mass and the non-interacting meson-meson energy threshold. One may infer from Sec.~\ref{sec:theory} that when the value of threshold energy, $E$, is far from the $b\bar b$ mass, $M$, the coupling effects are small. Moreover, when $M<E$ the effective potential is repulsive and it is unlikely that the coupling can generate relevant non-perturbative emergent phenomena; in fact, one usually obtains the same but dressed state due to the influence of near thresholds, moving to lower masses. However, if $M>E$ the effective potential becomes negative and a variety of emergent phenomena such as dynamically generated states with a dominant molecular structure may appear.

\begin{table}[!t]
\caption{\label{tab:Res1} Masses and widths, in MeV, of the poles predicted by our constituent-quark-model-based meson-meson coupled-channels calculation. Theoretical uncertainties have been estimated by modifying the most relevant model parameters within a range of 10\%.}
\begin{ruledtabular}
\begin{tabular}{crr}
State & $M_{\rm pole}$ & $\Gamma_{\rm pole}$ \\
\hline
1 & $10562 \pm 1$ & $29 \pm 5$ \\
2 & $10601 \pm 5$ & $3 \pm 2$ \\
3 & $10645 \pm 1$ & $23 \pm 1$ \\
4 & $10694 \pm 1$ & $8 \pm 1$ \\
5 & $10712 \pm 5$ & $41 \pm 4$ \\
6 & $10835 \pm 2$ & $52 \pm 7$ \\
7 & $10859 \pm 2$ & $13 \pm 3$ \\
8 & $10888 \pm 1$ & $3 \pm 1$ \\
\end{tabular}
\end{ruledtabular}
\end{table}

\begin{figure}[!t]
\centerline{\includegraphics[width=0.45\textwidth]{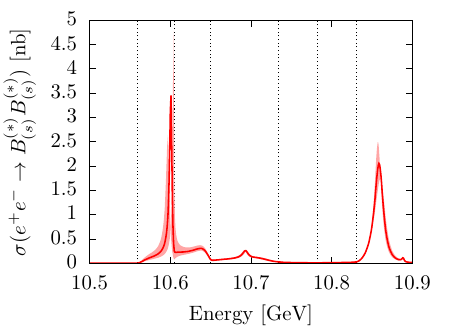}}
\caption{\label{fig:epempro} Production cross section, in nb, of the $\Upsilon$ states when annihilating a pair of electron-positron and measuring in the final state one of the $B_{(s)}^{(\ast)}\bar B_{(s)}^{(\ast)}$ channels. The vertical black dashed lines indicate the open-bottom meson-meson thresholds which, from left to right, are $B\bar B$, $B\bar B^\ast$, $B^\ast \bar B^\ast$, $B_s\bar B_s$, $B_s\bar B_s^\ast$ and $B_s^\ast \bar B_s^\ast$. Theoretical uncertainties have been estimated by modifying the most relevant model parameters within a range of 10\%.}
\end{figure}

\begin{table*}[!t]
\caption{\label{tab:Res2} Probability, in $\%$, of each channel that compose the wave function of the poles predicted by our constituent-quark-model-based meson-meson coupled-channels calculation. Theoretical uncertainties have been estimated by modifying the most relevant model parameters within a range of 10\%.}
\begin{ruledtabular}
\scalebox{0.90}{
\begin{tabular}{crrrrrrrrrr}
State & ${\cal P}_{\Upsilon(4S)}$ & ${\cal P}_{\Upsilon(3D)}$ & ${\cal P}_{\Upsilon(5S)}$ & ${\cal P}_{\Upsilon(4D)}$  & ${\cal P}_{B\bar B}$ & ${\cal P}_{B\bar B^*}$ & ${\cal P}_{B^*\bar B^*}$ & ${\cal P}_{B_s\bar B_s}$ & ${\cal P}_{B_s\bar B_s^*}$ & ${\cal P}_{B_s^*\bar B_s^*}$ \\
\hline
1  & $40_{-4}^{+3}$  & $16 \pm 5$  & $1.5_{-0.3}^{+0.4}$  & $1.5_{-0.5}^{+0.6}$   & $33 \pm 8$  & $1.9_{-0.5}^{+0.6}$  & $5.2_{-0.6}^{+0.4}$  & $0.141_{-0.006}^{+0.002}$  & $0.05 \pm 0.01$  & $0.24_{-0.03}^{+0.01}$  \\
2  & $19_{-5}^{+3}$  & $1.1 \pm 0.4$  & $0.13_{-0.04}^{+0.02}$  & $0.04 \pm 0.01$   & $48_{-2}^{+1}$  & $28_{-2}^{+6}$  & $4 \pm 1$  & $0.017_{-0.007}^{+0.005}$  & $0.11 \pm 0.04$  & $0.10 \pm 0.04$  \\
3  & $22_{-4}^{+5}$  & $9 \pm 1$  & $0.07 \pm 0.01$  & $0.14_{-0.02}^{+0.01}$   & $6 \pm 1$  & $34.1_{-0.6}^{+0.4}$  & $28_{-2}^{+1}$  & $0.14 \pm 0.02$  & $0.102_{-0.006}^{+0.004}$  & $0.193_{-0.009}^{+0.008}$  \\
4  & $1.7_{-0.2}^{+0.1}$  & $40 \pm 2$  & $0.027_{-0.004}^{+0.001}$  & $0.007_{-0.001}^{+0.002}$   & $12.0_{-0.7}^{+0.8}$  & $16 \pm 1$  & $29.6 \pm 0.1$  & $0.32_{-0.04}^{+0.05}$  & $0.32_{-0.06}^{+0.07}$  & $0.029_{-0.004}^{+0.005}$  \\
5  & $36 \pm 5$  & $8.8_{-1.0}^{+0.8}$  & $3.3 \pm 0.2$  & $0.166 \pm 0.003$   & $4.5_{-0.9}^{+1.0}$  & $2.74_{-0.04}^{+0.09}$  & $43_{-5}^{+4}$  & $0.9 \pm 0.2$  & $0.22_{-0.02}^{+0.03}$  & $0.55_{-0.07}^{+0.06}$  \\
6  & $4 \pm 1$  & $0.71_{-0.04}^{+0.06}$  & $80_{-3}^{+2}$  & $1.4 \pm 0.1$   & $1.3_{-0.3}^{+0.4}$  & $2.1_{-0.4}^{+0.5}$  & $6 \pm 1$  & $0.4 \pm 0.1$  & $1.2_{-0.3}^{+0.4}$  & $3 \pm 1$  \\
7  & $0.4_{-0.2}^{+0.3}$  & $0.03 \pm 0.01$  & $53.7_{-0.9}^{+0.0}$  & $0.2_{-0.1}^{+0.2}$   & $1.9 \pm 0.1$  & $4.4_{-0.1}^{+0.2}$  & $5.7 \pm 0.8$  & $1.33_{-0.04}^{+0.06}$  & $2.1 \pm 0.4$  & $30 \pm 1$  \\
8  & $0.004 \pm 0.002$  & $0.005 \pm 0.002$  & $0.05_{-0.02}^{+0.03}$  & $50.23_{-0.02}^{+0.00}$   & $16.6 \pm 0.1$  & $16.1_{-0.3}^{+0.2}$  & $2.80_{-0.04}^{+0.05}$  & $4.9 \pm 0.1$  & $8.5 \pm 0.2$  & $0.8_{-0.1}^{+0.2}$  \\
\end{tabular}}
\end{ruledtabular}
\end{table*}

\begin{table*}[!t]
\caption{\label{tab:Res3} Decay branching fractions, in $\%$, to the different open-bottom meson-meson channels of the poles predicted by our constituent-quark-model-based meson-meson coupled-channels calculation. Theoretical uncertainties have been estimated by modifying the most relevant model parameters within a range of 10\%.}
\begin{ruledtabular}
\begin{tabular}{crrrrrr}
State & ${\cal BR}_{B\bar B}$ & ${\cal BR}_{B\bar B^*}$ & ${\cal BR}_{B^*\bar B^*}$ & ${\cal BR}_{B_s\bar B_s}$ & ${\cal BR}_{B_s\bar B_s^*}$ & ${\cal BR}_{B_s^*\bar B_s^*}$ \\
\hline
1  & $100 \pm 0$  & $0 \pm 0$  & $0 \pm 0$  & $0 \pm 0$  & $0 \pm 0$  & $0 \pm 0$  \\
2  & $100 \pm 0$  & $0 \pm 0$  & $0 \pm 0$  & $0 \pm 0$  & $0 \pm 0$  & $0 \pm 0$  \\
3  & $6 \pm 1$  & $94 \pm 1$  & $0 \pm 0$  & $0 \pm 0$  & $0 \pm 0$  & $0 \pm 0$  \\
4  & $23 \pm 1$  & $14_{-2}^{+1}$  & $63.2_{-0.2}^{+0.1}$  & $0 \pm 0$  & $0 \pm 0$  & $0 \pm 0$  \\
5  & $8_{-1}^{+2}$  & $3.0_{-0.2}^{+0.4}$  & $90 \pm 1$  & $0 \pm 0$  & $0 \pm 0$  & $0 \pm 0$  \\
6  & $7.4_{-0.3}^{+0.1}$  & $4.3_{-0.1}^{+0.3}$  & $5.5_{-0.1}^{+0.6}$  & $2 \pm 0$  & $4 \pm 1$  & $77_{-2}^{+0}$  \\
7  & $4.1_{-0.4}^{+0.3}$  & $8 \pm 0$  & $2.2_{-0.2}^{+0.1}$  & $3.0 \pm 0.1$  & $1.8_{-0.4}^{+0.6}$  & $81.4_{-0.2}^{+0.1}$  \\
8  & $33.9_{-0.3}^{+0.4}$  & $32.9_{-0.5}^{+0.3}$  & $5 \pm 0$  & $9.8_{-0.2}^{+0.3}$  & $17.3 \pm 0.3$  & $0.8 \pm 0.1$  \\
\end{tabular}
\end{ruledtabular}
\end{table*}

We perform a coupled-channels calculation in which the bare states $\Upsilon(4S)$, $\Upsilon(3D)$, $\Upsilon(5S)$ and $\Upsilon(4D)$ are considered together with the threshold channels $B\bar{B}$, $B\bar{B}^\ast$, $B^\ast \bar{B}^\ast$, $B_s\bar{B}_s$, $B_s\bar{B}_s^\ast$ and $B_s^\ast \bar{B}_s^\ast$. In order to compensate for the large mass-shifts that higher thresholds add to the bare states, which are already coded in the screened confinement potential, we move up the masses of the bare states $35$ MeV. Our results are shown in Tables~\ref{tab:Res1},~\ref{tab:Res2} and~\ref{tab:Res3}. First mentioned table shows the pole position in complex energy plane characterized by mass and width ($E=M-i\,\Gamma/2$), the second one gives the probability of each channel that compose the wave function of the dressed hadron and the third table provides hadron's decay branching fractions to the different open-bottom meson-meson channels considered herein.

Concerning Table~\ref{tab:Res1}, eight poles in the complex energy plane are predicted; all of them are resonances, \emph{i.e.} they are singularities that appear in the physical sheet. It may seem like there are many, al least more than observed experimentally. However, as shown in Fig.~\ref{fig:epempro}, only three peaks are present in the most common production process of $\Upsilon$ states: $e^+e^-\to B_{(s)}^{(\ast)}\bar B_{(s)}^{(\ast)}$. The two dominant peaks appear at around the masses of well-established $\Upsilon(4S)$ and $\Upsilon(10860)$ states, the small bump at approximately $10.7\,\text{GeV}$ is our assignment of the $\Upsilon(10750)$ signal observed in $e^+e^-\to \pi^+\pi^-\Upsilon(nS)$. If our interpretation is correct, we have learned from Table~\ref{tab:Res1} and Fig~\ref{fig:epempro} that (i) a richer complex spectrum is gained when thresholds are present and bare bound states are sufficiently non-relativistic; (ii) those poles obtained in the complex energy plane do not have to appear as simple peaks in the relevant cross sections due to many reasons such as their distance from the energy real-axis, coupling with the corresponding final channel, etc; and (iii) when comparing with experiment, the so-called $\Upsilon(4S)$ and $\Upsilon(10860)$ signals are clearly identified but there should be another small one, corresponding to the $\Upsilon(10750)$ case, whose origin cannot be traced back to any bare quark-antiquark bound state.

It is convenient to analyze Tables~\ref{tab:Res2} and~\ref{tab:Res3} together. The first pole has a mass, $10562\,\text{MeV}$, and a total decay width, $29\,\text{MeV}$, compatible with those values collected in the PDG for the $\Upsilon(4S)$ state. Moreover, its wave function has as the dominant channel the $\Upsilon(4S)$, \emph{i.e.} a canonical bottom-antibottom 4S bound state; followed by $B\bar B$ and $\Upsilon(3D)$ components; moreover, its decay branching fraction of $B\bar B$ is $100\%$. Therefore, the natural assignment to this dressed state is the experimentally observed $\Upsilon(4S)$ state.

The following two poles seems to be singularities produced dynamically in the complex energy plane due to the coupling between conventional bottomonium states and meson-meson thresholds. In fact this coupling makes them mostly $B\bar B-B\bar B^\ast$ and $B\bar B^\ast-B^\ast \bar B^\ast$ molecules, respectively; having also measurable traces of $\Upsilon(4S)$ and $\Upsilon(3D)$ components in their wave functions. Nevertheless, as one can see in Fig.~\ref{fig:epempro}, these structures do not materialize in the production cross section.

We assign the fourth pole to the so-called $\Upsilon(3D)$ state. Our predictions for the mass and total decay width are $10645\,\text{MeV}$ and $23\,\text{MeV}$. Its wave function exhibit a dominant $3D$ $b\bar b$ constituent, followed by important $B\bar B$, $B\bar B^\ast$ and $B^\ast \bar B^\ast$ components. Table~\ref{tab:Res3} shows that the decay branching fractions of this state into the $B\bar B$, $B\bar B^\ast$ and $B^\ast \bar B^\ast$ final channels are $23\%$, $14\%$ and $63\%$, respectively. As one can see in Fig.~\ref{fig:epempro}, this state takes part on the second bump observed in production cross section; however, its contribution is small because it depends on the value of $I_4$ in Eq.~\eqref{eq:I4integral}, which is small compared to the contribution of $S$-wave $\Upsilon$ states.

The last observation is connected with our interpretation of the fifth pole as the $\Upsilon(10750)$ candidate because, as one can see in Table~\ref{tab:Res2}, its corresponding wave function shows a large $\Upsilon(4S)$ component which provides the leverage for its production. Note also that the wavefunction's $B^\ast \bar B^\ast$ channel is of the same order of magnitude than the former, encouraging us to conclude that this dressed hadronic state is in fact a resonance whose structure is an equally mixture of a conventional $b\bar b$ $4S$ state and $B^\ast \bar B^\ast$ molecule. Looking at Table~\ref{tab:Res1}, its theoretical mass and width are $10712\,\text{MeV}$ and $41\,\text{MeV}$, respectively, which are in reasonable agreement with the experimental data, Eq.~\eqref{eq:experiment}. Finally, Table~\ref{tab:Res3} shows that this state decays $90\%$ of the time into $B^\ast \bar B^\ast$ followed by the $B\bar B$ final state with a branching fraction of $8\%$.

Concerning the bump that can be seen in the production cross section (Fig.~\ref{fig:epempro}) at around $10.85\,\text{GeV}$, it is mostly produced by the sixth pole because, as shown in Table~\ref{tab:Res2}, its wave function exhibits a very dominant $5S$ $b\bar b$ component, with a probability of $80\%$; being the rest channels an order of magnitude less probable. Without changing our attention from Table~\ref{tab:Res2}, the other two poles are constituted by an equally mixture of canonical bottomonium structure and open-bottom meson-meson molecule. Table~\ref{tab:Res3} shows the branching fractions of these three dressed states to $B_{(s)}^{(\ast)}\bar B_{(s)}^{(\ast)}$ channels. For our tentative assignment of the $\Upsilon(5S)$ state as $\Upsilon(10860)$, the theoretical mass and width are in fair agreement with experiment; our mass value is a bit lower than the experimental one, $10885\,\text{MeV}$, whereas our width is higher than the PDG's figure, $37\,\text{MeV}$. This may be explained by the projection in real-axis of the singularities shown in the complex energy plane at the relevant energy, which is the only measurable feature. It is also worth to note that we provide correct orders of magnitude for the branching fractions of the $B^{(\ast)}\bar B^{(\ast)}$ state whereas unfortunately the one corresponding to the $B_s^\ast \bar B_s^\ast$ channel is far larger than experiment.


\section{Summary}
\label{sec:summary}

The $\Upsilon(10753)$ signal seems to be a potential candidate of the $\Upsilon$-family. It was firstly observed by the Belle collaboration in 2019 and, later on, by the Belle~II collaboration in 2022. The joint significance is $5.2$ standard deviations; moreover, its mass, total decay width and quantum numbers are also determined: $M=10753\,\text{MeV}$, $\Gamma=36_{-12}^{+18}\,\text{MeV}$ and $J^P=1^-$. Since this new state does not fit into the spectrum predicted by any reasonable constituent quark model, everything points that it may be an exotic state whose nature is explained by some mechanism that goes beyond the simple quark-antiquark interaction. The most logical extension to the naive quark model is the coupling of bare bottomonia with their closest open-bottom thresholds in order to assess if emergent non-perturbative dynamical states could be produced.

We have analyzed the predicted spectrum of the $\Upsilon$-family, within an energy range around the $\Upsilon(10753)$'s mass, in the framework of a constituent quark model~\cite{Vijande:2004he} which satisfactorily describes a wide range of properties of conventional hadrons containing heavy quarks~\cite{Segovia:2013wma, Segovia:2016xqb}. The quark-antiquark and meson-meson degrees of freedom have been incorporated with the goal of elucidating the influence of open-bottom meson-meson thresholds into the concerning bare states and to shed some light on the nature and structure of the $\Upsilon(10753)$ state. In particular, we have performed a coupled-channels calculation in which the bare states $\Upsilon(4S)$, $\Upsilon(3D)$, $\Upsilon(5S)$ and $\Upsilon(4D)$ are considered together with the threshold channels $B\bar{B}$, $B\bar{B}^\ast$, $B^\ast \bar{B}^\ast$, $B_s\bar{B}_s$, $B_s\bar{B}_s^\ast$ and $B_s^\ast \bar{B}_s^\ast$. 

Among the results we have described, the following conclusions are of particular interest: (i) a richer complex spectrum is gained when thresholds are present and bare bound states are sufficiently non-relativistic; (ii) those poles obtained in the complex energy plane do not have to appear as simple peaks in the relevant cross sections; and (iii) the $\Upsilon(10750)$ candidate is interpreted as a dressed hadronic resonance whose structure is an equally mixture of a conventional $b\bar b$ state and $B^\ast B^\ast$ molecule.


\begin{acknowledgments}
Work partially financed by 
EU Horizon 2020 research and innovation program, STRONG-2020 project, under grant agreement no. 824093;
Ministerio Espa\~nol de Ciencia e Innovaci\'on under grant Nos. PID2019-105439GB-C22, PID2019-107844GB-C22 and PID2022-140440NB-C22;%
Junta de Andaluc\'ia under contract Nos. Operativo FEDER Andaluc\'ia 2014-2020 UHU-1264517, P18-FR-5057 and also PAIDI FQM-370.
\end{acknowledgments}


\bibliography{PrintUpsilonAtThreshold}

\end{document}